\documentclass[]{spie}  

\usepackage{amsmath,amsfonts,amssymb}
\usepackage[colorlinks=true, allcolors=blue]{hyperref}
\usepackage{csquotes}

\usepackage{graphicx,caption,subcaption}
\usepackage{enumitem}
\graphicspath{{plots/}}

\title{High contrast imaging of exoplanets on ELTs using a super-Nyquist wavefront control scheme}

\author[a,b]{Benjamin L. Gerard}
\author[b,a]{Christian Marois}
\affil[a]{University of Victoria, Department of Physics and Astronomy, 3800 Finnerty Rd., Victoria, V8P 5C2, Canada}
\affil[b]{National Research Council of Canada, 5071 West Saanich Rd, Victoria, V9E 2E7, Canada}

\authorinfo{Further author information: (Send correspondence to B. Gerard)\\B. Gerard.: E-mail: bgerard@uvic.ca}

\begin{document} 
\maketitle

\begin{abstract}
One of the key science goals for extremely large telescopes (ELTs) is the detailed characterization of already known directly imaged exoplanets. The typical adaptive optics (AO) Nyquist control region for ELTs is $\sim$0.4 arcseconds, placing many already known directly imaged planets outside the DM control region and not allowing any standard wavefront control scheme to remove speckles that would allow higher SNR images/spectra to be acquired. This can be fixed with super-Nyquist wavefront control (SNWFC), using a sine wave phase plate to allow for wavefront control outside the central DM Nyquist region. We demonstrate that SNWFC is feasible through a simple, deterministic, non-coronagraphic, super-Nyquist speckle nulling technique in the adaptive optics laboratory at the National Research Council of Canada. We also present results in simulation of how SNWFC using the self coherent camera (SCC) can be used for high contrast imaging. This technique could be implemented on future high contrast imaging instruments to improve contrast outside the standard central dark hole for higher SNR characterization of exoplanets.
\end{abstract}


\section{INTRODUCTION}
\label{sec:intro} 

The direct imaging of exoplanets is more sensitive to planets beyond $\sim$5-10 AU. Although direct imaging has seen less planet detections than radial velocity or transit techniques, the past eight years have revealed a handful of directly imaged planets, including multiple planets around HR 8799 \cite{8799_1, 8799_2}, HD 95086 b\cite{95086}, Beta Pic b\cite{beta_pic}, and most recently 51 Eri b\cite{51eri}. 

Detailed characterization of these existing planets with extremely large telescopes (ELTs) may be a difficult task, since these systems may lie either at the edge of or outside of the typical $\sim$0.4 arcsecond ELT adaptive optics (AO) Nyquist control region when observing in the near infrared\cite{nfiraos}. This region is set by the deformable mirror (DM) actuator pitch projected onto the telescope pupil, and for a square grid DM is a $(N_\text{act})(\lambda/D)\times(N_\text{act})(\lambda/D)$ region around the on-axis PSF, where $N_\text{act}$ is the number of actuators in width across the telescope pupil, $\lambda$ is the wavelength of light, and $D$ is the telescope diameter\cite{snwfc}. Thus, with classical, single conjugate AO (SCAO), uncorrected atmospheric turbulence and and quasi-static speckles will lower the planet signal to noise ratio (SNR).

A recent technique has recently been proposed to allow wavefront control outside the Nyquist control region, called super-Nyquist wavefront control (SNWFC)\cite{snwfc}. The main hardware component in this technique requires the use of a super-Nyquist element in an AO system, such as a mild pupil plane diffraction grating with a spacing between lines that is smaller than the DM actuator pitch relative to the pupil size. The pupil plane imprint creates a PSF copy in the focal plane that is outside the DM Nyquist region, allowing wavefront control to work in a similar $(N_\text{act})(\lambda/D)\times(N_\text{act})(\lambda/D)$ control region around this super-Nyquist PSF copy.

In this paper, we present the results of a laboratory experiment and simulations for a future experiment to show that it is possible to use SNWFC on an ELT AO system system to directly image already known and new exoplanets. In \S\ref{lab} we describe the laboratory experiment design (\S\ref{lab_design}), simulations of expected lab performance (\S\ref{lab_sims}), and results in the lab (\S\ref{lab_results}). In \S\ref{scc} we describe the setup and results of our simulation using the self coherent camera (SCC)\cite{scc1,scc2}. In \S\ref{conclusion}, we summarize our results and discuss future work.

\section{DETERMINISTIC LABORATORY EXPERIMENT}
\label{lab}

We first design, simulate, and test a simple deterministic laboratory experiment that demonstrates the possible performance improvement using SNWFC. The goal here is to show that performance improvement is possible in an idealized case to motivate more realistic SNWFC simulations (\S \ref{scc}), the future testing of this technique in a laboratory, and the future ELT applications for high contrast imaging of exoplanets.

\subsection{Experiment Design}
\label{lab_design}

\begin{figure} [h]
\begin{center}
\begin{tabular}{c} 
\includegraphics[height=7cm]{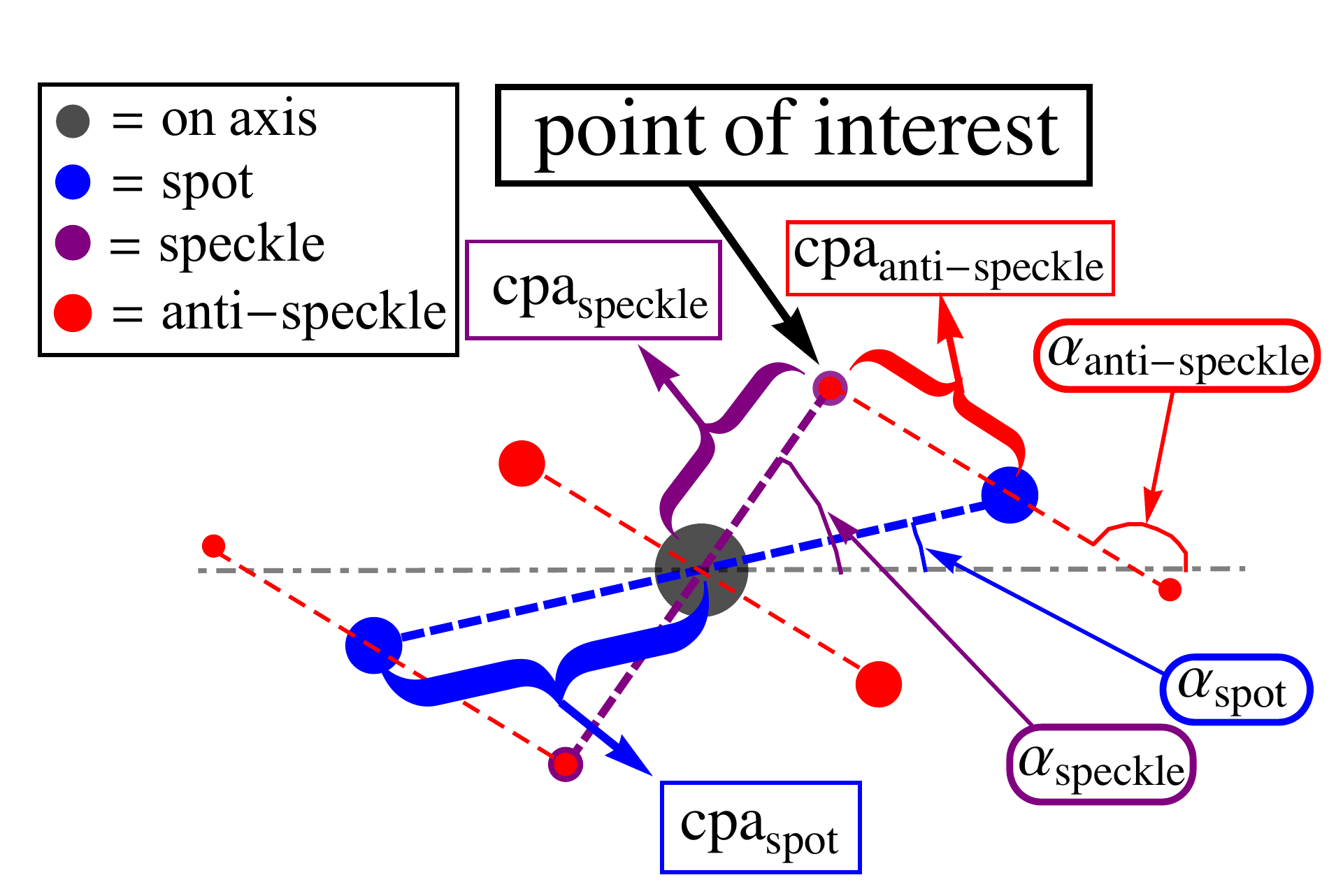}
\end{tabular}
\end{center}
\caption{A schematic diagram of the focal plane for our deterministic SNWFC laboratory experiment. \enquote{$\alpha$} represents the sine wave position angle, counter clockwise from the +$x$ axis. \enquote{cpa} is the number of cycles per aperture for a given DM sine wave, and it represents radial location of the sine wave PSF copy in the focal plane in units of $\lambda/D$ (the DM sine wave angular frequency is $f=2\pi(\text{cpa})/(D_\text{DM})$,where $D_\text{DM}$ is the DM pupil plane diameter). }
\label{fig: experiment_diagram} 
\end{figure}

In this section we describe the methodology and algorithm setup for our deterministic speckle nulling experiment. Figure \ref{fig: experiment_diagram} shows a schematic layout for our SNWFC experiment.

The basic structure involves the placement of three sine waves on the DM:
\begin{itemize}
\item \textbf{spot}: Shown in blue in Figure \ref{fig: experiment_diagram}, the spot represents the effect of a PSF copy from a pupil plane sine wave phase plate at a higher frequency than the DM Nyquist frequency, thus creating a copy of the on-axis PSF outside the AO control region. In the absence of a phase plate for our experiment, we use the highest DM sine wave frequency to represent the spot. We note that with this design there is no actual super-Nyquist element, since the spot is, by definition, sub-Nyquist. However, one can instead imagine a system with a lower order DM where the DM Nyquist frequency creates the speckle, in which case a super-Nyquist phase plate creates the spot. In this case, SNWFC is needed to null the speckle using the spot. But, the super-Nyquist phase plate for a lower order DM and our real DM spot sine wave have the same optical effect, and so for the purposes of this experiment we are still demonstrating SNWFC, but instead using less than the full DM to represent the Nyquist region.
\item \textbf{speckle}: Shown in purple in Figure \ref{fig: experiment_diagram}, the speckle is at a slightly lower frequency than the spot, again sub-Nyquist in this experiment, and is meant to represent speckle noise hiding the signal from a planet at the \textbf{point of interest} (Figure \ref{fig: experiment_diagram}).
\item \textbf{anti-speckle}: Shown in red in Figure \ref{fig: experiment_diagram}, the goal of this experiment is to copy the anti-speckle sine wave from the spot so that electric field in the focal plane at the point of interest is minimized, allowing the off axis planet light at that point to then be seen. Unlike the speckle intensity, the planet light is not removed because it is incoherent with the stellar light.
\end{itemize}

The frequency, position angle, amplitude, and phase of the spot and speckle are user-defined parameters. In order to cancel the speckle at the point of interest, the anti-speckle frequency and position angle are already determined geometrically based on the user-defined spot and speckle parameters. The phase and amplitude of the anti-speckle are then determined iteratively using the following methodology and procedure:

\begin{enumerate}

\item Using the image plane intensity, which is approximately a measure of amplitude squared, the anti-speckle amplitude is then approximately 
\begin{equation}
a_{\text{anti-speckle}}=(f) a_{\text{speckle}}\sqrt{\frac{\text{im}_{\text{spot}}}{\text{im}_{\text{speckle}}}}
\label{eq: fudge}
\end{equation}
in an image, im, with only two sine waves, the spot and speckle, where $a$ is the sine wave amplitude. The fudge parameter, $f$ is equal to 1 in this step and explained further below in step 3.

\item The anti-speckle phase is unknown, so we simply loop through -$\pi$ to $\pi $ in phase on the anti-speckle in order to find the phase that minimizes image plane intensity at the point of interest. 

\item Using $f=1$ in equation \ref{eq: fudge} is only a rough estimate for $a_\text{anti-speckle}$ because we are using a linear approximation of the wavefront ($\text{wavefront}=a\; e^{i\; \phi}\sim a(1+i\phi)$, so $|\text{wavefront}|^2\sim O(a^2)$) to null the speckle, neglecting the higher order terms. So, to correct for these higher order terms we add a fudge factor as necessary in the anti-speckle amplitude, where $0.5\lesssim f\lesssim$2. 

\end{enumerate}

In simulations (\S\ref{lab_sims}) and in the lab (\S\ref{lab_results}), finding a precise optimal anti-speckle phase is done iteratively, first finding a rough estimate, and then using a finer grid spacing in phase around that rough estimate to get a more precise value. The same iterative procedure is applied in finding the amplitude fudge factor. Throughout the remainder of \S\ref{lab_design}, we calculate contrast in a given image normalized to the peak value in that same image, since we do not use a coronagraph in this laboratory experiment.

\subsection{Simulations}
\label{lab_sims}

We ran simulations of two different sine wave configurations: one where the point of interest lies on a bright Airy ring of the on-axis PSF, and one where the point of interest lies on a nearby dark ring. The initial and final parameters for both simulations are shown in Table \ref{tab: parameters}. We use an image size of 2048$\times$2048 pixels, beam ratio of 4, and a circular greyscale pupil (instead of a binary mask, the pupil mask is mean binned 10$\times$10 from an original binary mask of image size of 20480$\times$20480 to prevent numerical pixelation effects).

\begin{table}[ht]
\caption{Initial simulation and final parameters for our three sine wave deterministic speckle nulling procedure for both our dark ring and Airy ring simulations. The spot and speckle parameters are set initially, while the anti-speckle phase and amplitude are determined iteratively via the focal plane wavefront sensing scheme described in \S\ref{lab_design}.} 
\label{tab: parameters}
\begin{center}       
\begin{tabular}{|l|l|l|l|l|}
\hline
\rule[-1ex]{0pt}{3.5ex}  \textbf{sine wave} & \textbf{$\alpha$ (degrees)} & \textbf{cpa} & \textbf{phase (radians)} & \textbf{amplitude (radians)} \\
\hline
\multicolumn{5}{|c|}{Airy ring simulation} \\
\hline
\rule[-1ex]{0pt}{3.5ex}  spot & 1.0 & 4.0 & 0 & 0.13  \\
\hline
\rule[-1ex]{0pt}{3.5ex}  speckle & 28.7 & 2.95 & 0 & 0.064  \\
\hline
\rule[-1ex]{0pt}{3.5ex}  anti-speckle & 135.0 & 2.0 & -2.57 & 0.94  \\
\hline
\multicolumn{5}{|c|}{dark ring simulation} \\
\hline
\rule[-1ex]{0pt}{3.5ex}  spot & 1.0 & 4.0 & 0 & 0.13  \\
\hline
\rule[-1ex]{0pt}{3.5ex}  speckle & 35.0 & 2.57 & 0 & 0.064  \\
\hline
\rule[-1ex]{0pt}{3.5ex}  anti-speckle & 142.0 & 2.4 & 0.84 & 0.85  \\
\hline
\end{tabular}
\end{center}
\end{table}

\begin{figure}[!h]
\begin{center}
\includegraphics[width=1.0\textwidth]{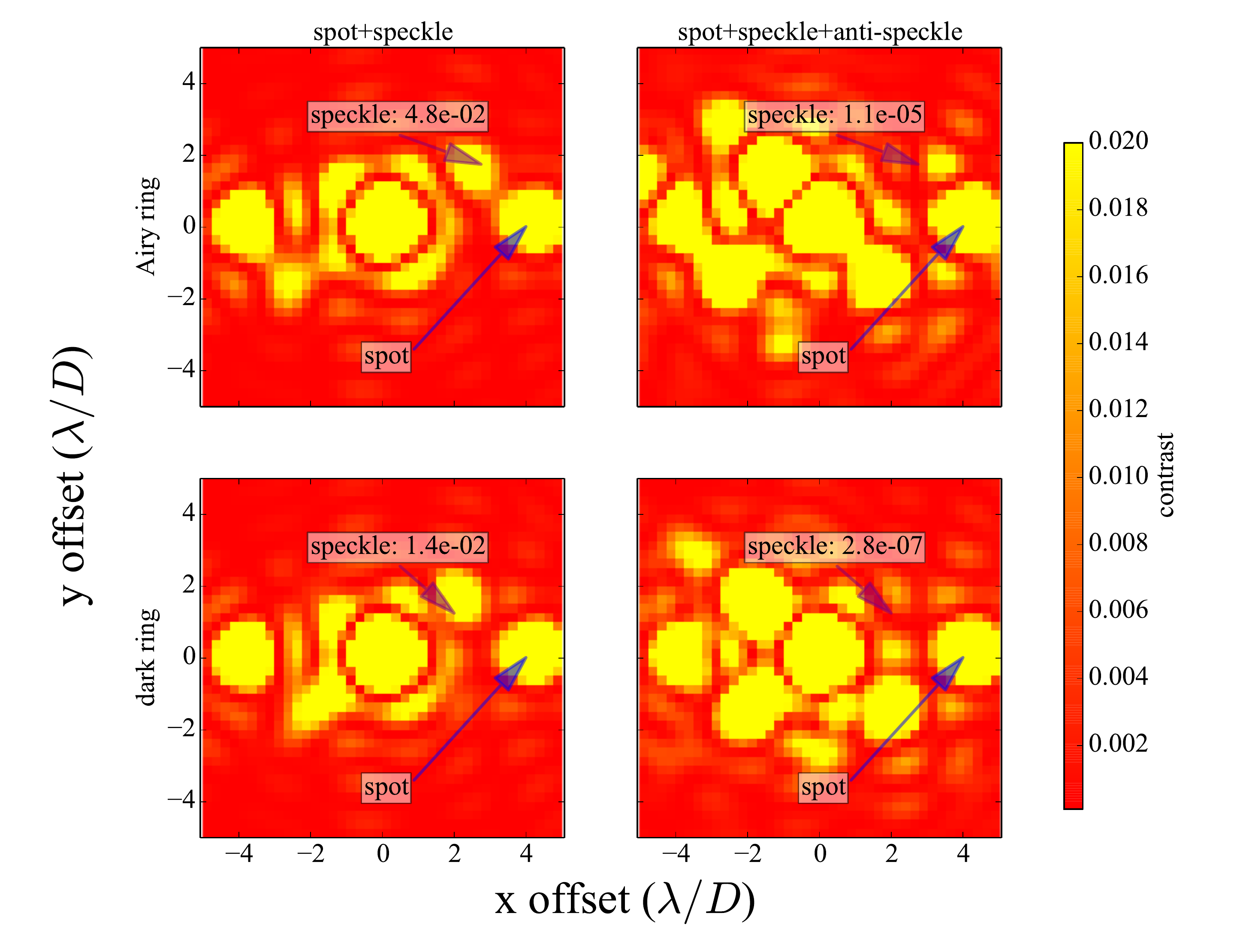}
\end{center}
\caption{Images of our Airy ring (first row) and dark ring (second row) deterministic speckle nulling simulations. The left column shows the initial configuration with only two sine waves (spot+speckle), and the right column shows the configuration with all three sine waves after reaching the minimum anti-speckle phase and amplitude.}
\label{fig: sim_images}
\end{figure}

The results of our simulations after anti-speckle phase and amplitude iterations are shown in Figure \ref{fig: sim_images}. Figure \ref{fig: a} shows the results after only phase iterations, to be compared later with our laboratory results in \S\ref{lab_results}. Our two simulations show that changing the point of interest from a bright Airy ring to a nearby dark ring changes the achievable contrast as well as the absolute anti-speckle phase. The final contrasts reached after anti-speckle amplitude iterations (equation \ref{eq: fudge}) for the Airy ring and dark ring simulations are $1.1\times10^{-5}$ and $2.8\times10^{-7}$, respectively. After amplitude iterations for the Airy ring simulation we found $f=1.786$, whereas for the dark ring simulation we found $f=1.004$. The fudge factor results are expected, since higher order terms in the Taylor expansion of the PSF fall off more quickly in the dark ring compared to the Airy ring\cite{perrin_psf}.

\subsection{Laboratory Results}
\label{lab_results}

\begin{figure}[!h]
\begin{center}
\includegraphics[width=1.0\textwidth]{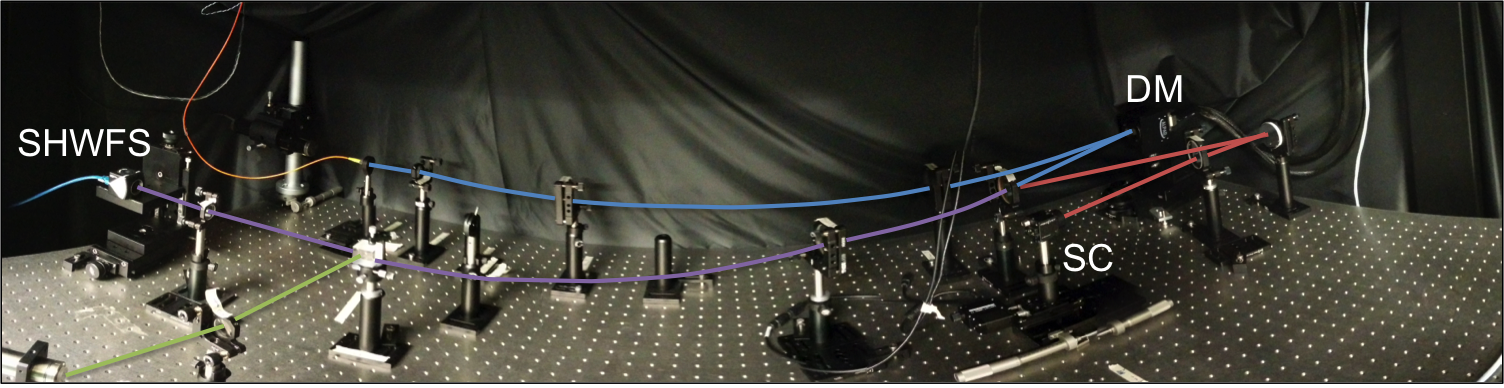}
\end{center}
\caption{The NRC Adaptive Optics Laboratory wavefront sensor bench. The common path is in blue, the SHWFS path is in purple, the non-common path to the science camera (SC) is in red, and a separate pyramid WFS path (not fully shown) is in green.}
\label{fig: bench_pic}
\end{figure}

We use a simple SCAO setup in the Adaptive Optics Laboratory at the National Research Council, Astronomy and Astrophysics (NRC), shown in Figure \ref{fig: bench_pic}. We use a monochromatic 655 nm fiber-fed laser diode as a light source, an $11\times11$ actuator ALPAO DM (circular pupil, 97 total actuators), a $32\times40$ subaperture Imagine Optic HASO Shack Hartman wavefront sensor (SHWFS), ultimately using a 29 subaperture diameter to remove edge effects. 

We apply sine waves to the DM in a closed loop procedure, using the expected SHWFS slopes for the desired sine wave amplitude, frequency, and phase, described below. The $x$ and $y$ SHWFS slopes for a DM sine wave at a subaperture, $i$ (using 1 indexing), are the mean values of the derivative of the wavefront phase along $x$ and $y$, respectively:
\begin{align}
\label{eq: shwfs_slopes}
x_\text{slope}(i)&=\left\langle \frac{\partial \Phi }{\partial x}\right\rangle_i \nonumber \\
&= \frac{\int _{x_i-d/2}^{x_i+d/2}\int _{y_i-d/2}^{y_i+d/2}\frac{\partial \Phi }{\partial x}d x d y}{\int _{x_i-d/2}^{x_i+d/2}\int _{y_i-d/2}^{y_i+d/2}d
x d y}, \\
y_\text{slope}(i)&=\left\langle \frac{\partial \Phi }{\partial y}\right\rangle_i \nonumber \\
&= \frac{\int _{x_i-d/2}^{x_i+d/2}\int _{y_i-d/2}^{y_i+d/2}\frac{\partial \Phi }{\partial y}d x d y}{\int _{x_i-d/2}^{x_i+d/2}\int _{y_i-d/2}^{y_i+d/2}dx d y}, \nonumber
\end{align}
where $\Phi$ is the wavefront phase, $d$ is the size of one subaperture ($d=D_\text{pup}/n_\text{subapertures}$, where $D_\text{pup}$ and $n_\text{subapertures}$ are the diameter and number of subapertures across the SHWFS pupil, respectively), and $(x_i,y_i)$ is the center of subaperture $i$ in units of pupil diameter\footnote{Using this coordinate system, the lower left corner of the lower left subaperture in use is $(x_1,y_1)=(0,0)$ and the upper right corner of the upper right subaperture is $(x_{n_\text{subapertures}},y_{n_\text{subapertures}})=(D_\text{pup},D_\text{pup})$.}. After calibrating a flat wavefront with the reference slopes from closing the loop on zeros, a sine wave on the DM imparts a wavefront phase shift of
\begin{equation}
\Phi= a\; \text{sin}\left(\frac{2\pi}{D_\text{pup}}\text{cpa}(x\; \text{cos}(\alpha)-y\; \text{sin}(\alpha))-\phi\right),
\label{eq: sine}
\end{equation}
where $a,\; \alpha,\; \text{and } \phi$ are the sine wave the amplitude, position angle, and phase, respectively. Combining equations \ref{eq: shwfs_slopes} and \ref{eq: sine} allows us to apply a sine wave on the DM in closed loop at any amplitude, frequency (sub-Nyquist), position angle, and phase. 

\begin{figure} [!h]
\begin{center}
\includegraphics[height=8cm]{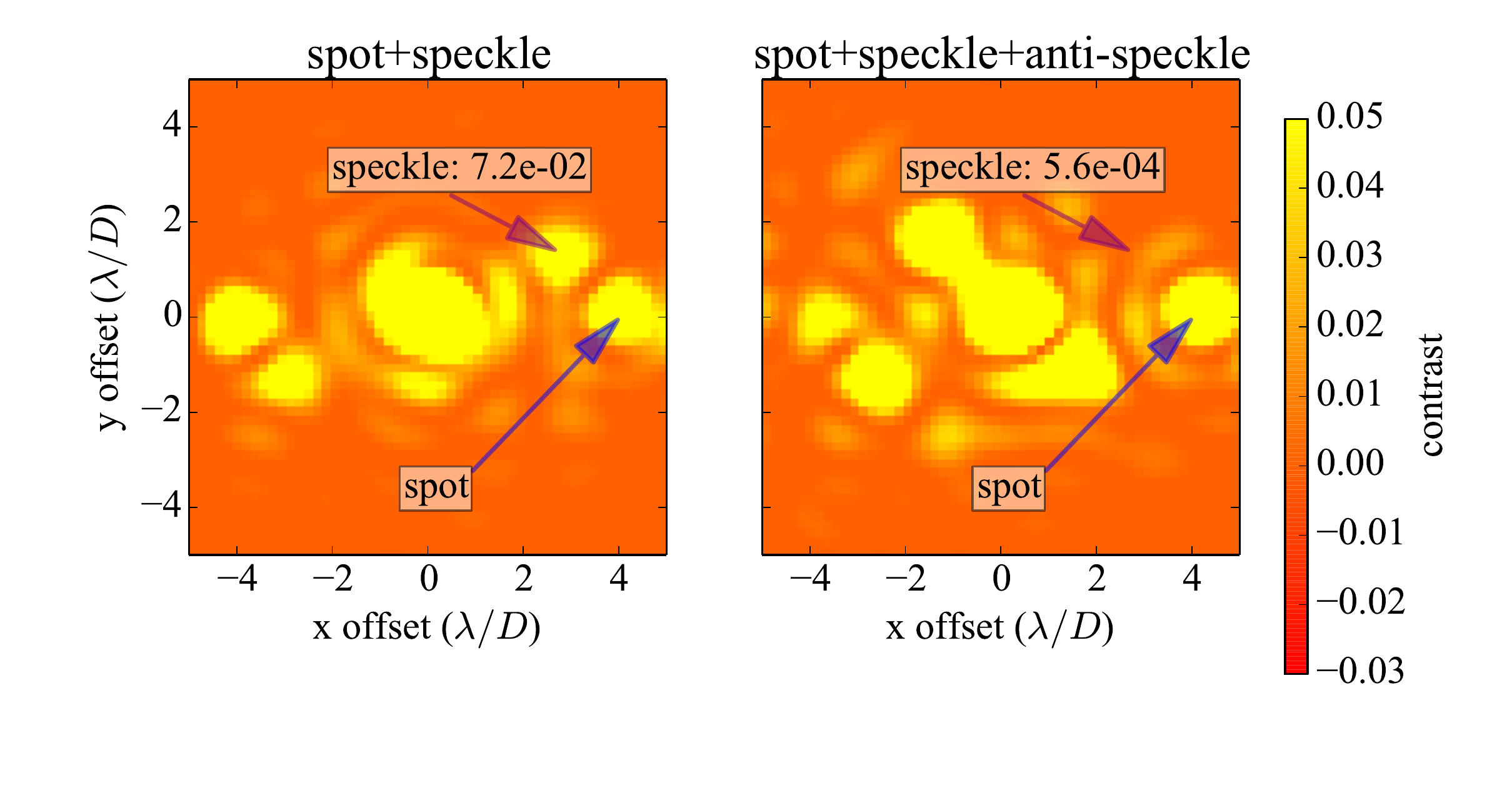}
\end{center}
\caption{ \label{fig: bench} 
Laboratory science camera images from our deterministic SNWFC experiment, using the Airy ring simulation parameters from Table \ref{tab: parameters}. The left image shows the initial speckle contrast for the spot+speckle DM sine wave combination, and the right image shows the spot+speckle+anti-speckle combination at the minimum anti-speckle phase of $\phi=-0.12$. The images are background/dark-subtracted, hence the negative contrast scale.}
\end{figure} 
\begin{figure} [!h]
\begin{center}
	\begin{subfigure}[b]{0.45\textwidth}
		\includegraphics[width=1.0\textwidth]{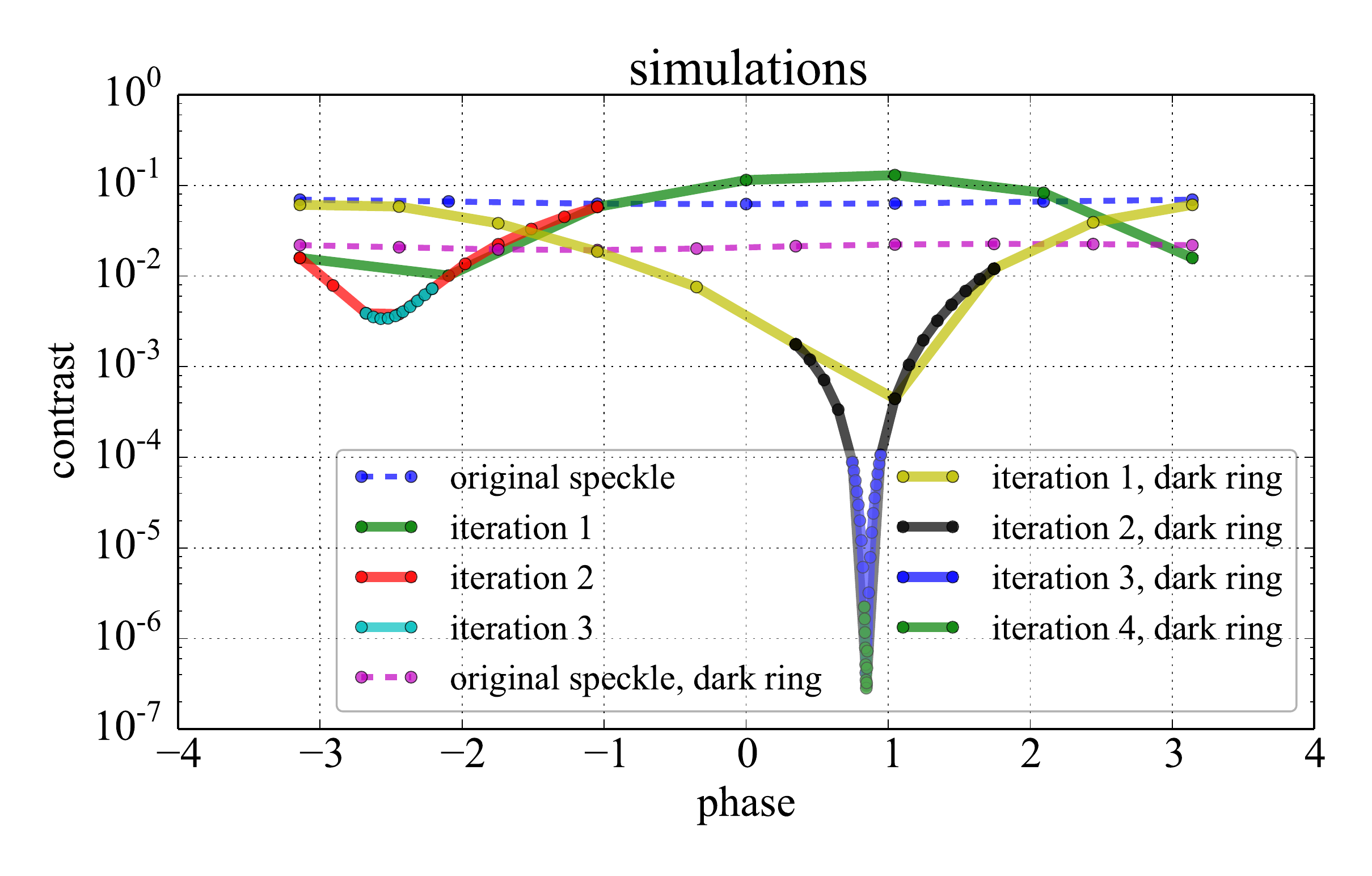}
		\caption{}
		\label{fig: a}
	\end{subfigure}
	\begin{subfigure}[b]{0.45\textwidth}
		\includegraphics[width=1.0\textwidth]{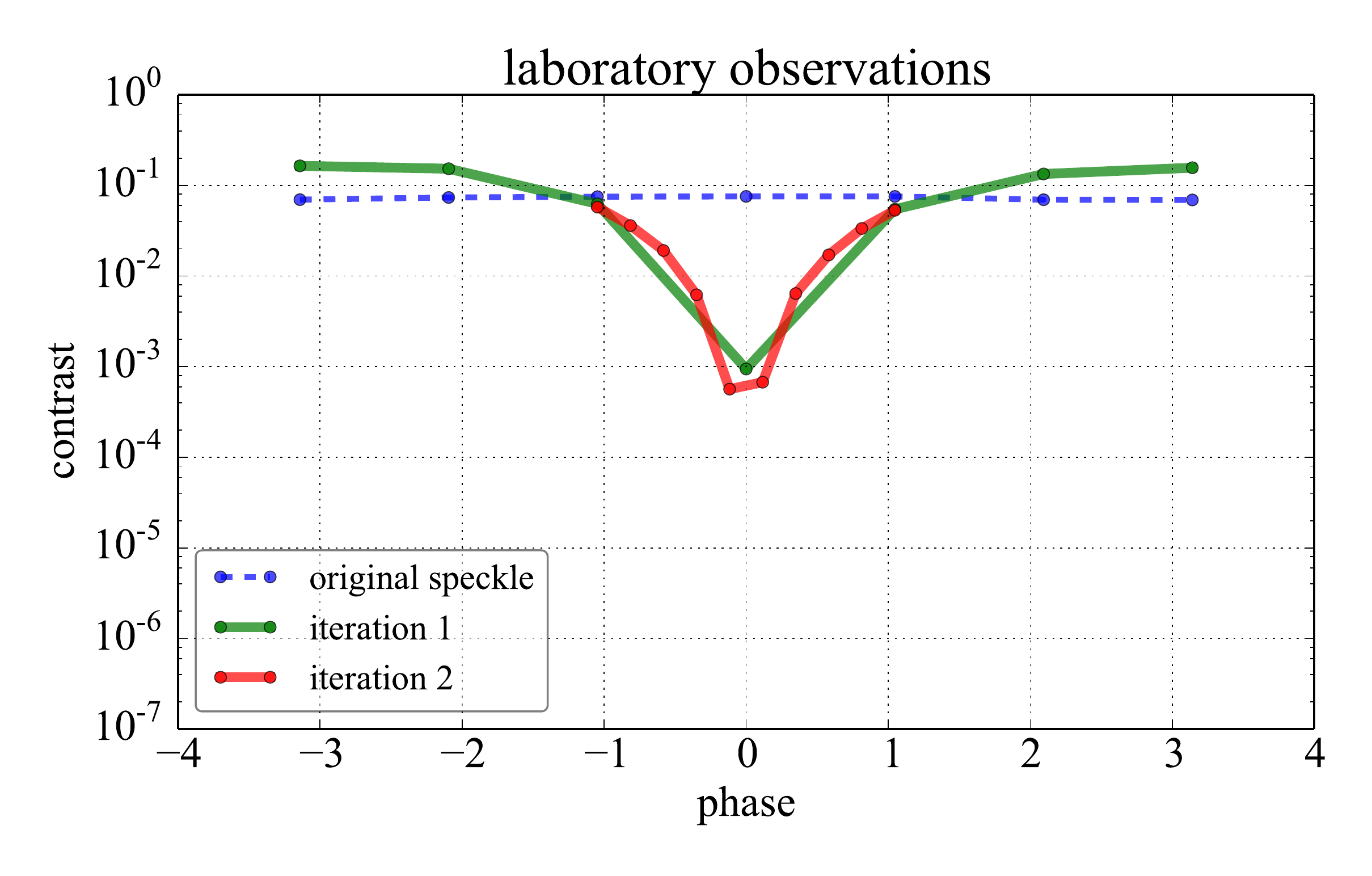}
		\caption{}
		\label{fig: b}
	\end{subfigure}
\end{center}
\caption{Speckle contrast plotted vs. anti-speckle phase (radians) at the point of interest. (a) Noiseless simulations for our SNWFC laboratory experiment, showing both the replicated laboratory setup, where the speckle location lies on an Airy ring, and a second simulation showing the effect of moving the speckle location into a nearby dark ring. (b) Results from our laboratory data. The minimum contrast and phase are both in between our two simulations in (a), suggesting that noise on the bench is moving the speckle location closer to a dark ring from the original Airy ring location.}
\label{fig: experiment_sims}
\end{figure} 
Using this methodology and the same parameters as our Airy ring simulation in Table \ref{tab: parameters} (but not the anti-speckle phase and amplitude, which we again determine iteratively), we implemented the deterministic speckle nulling procedure described in \ref{lab_design}, but after two phase iterations we reached the SC detector noise floor at the point of interest, and so we did not do any amplitude iterations. The before and after images comparing the spot+speckle with the minimum spot+speckle+anti-speckle are shown in Figure \ref{fig: bench}, and our laboratory phase iterations are shown in Figure \ref{fig: b}.

We found a final contrast at the point of interest of $5.6\times10^{-4}$, using $a_\text{anti-speckle}=88$ nm and $\phi_\text{anti-speckle}=-0.12$ (from equations \ref{eq: fudge} and \ref{eq: sine}, respectively). 

Comparing the final anti-speckle phase and contrast from our lab results in Figure \ref{fig: b} to our simulation results in Figure \ref{fig: a} suggests that in the lab, the speckle lies somewhere in between a bright and dark ring. Despite our laboratory setup identical to the Airy ring simulation, we suspect that noise on the bench distorts the on-axis Airy ring locations so that the speckle is located near a dark ring, suggested by the deeper contrast and closer absolute phase shift for the dark ring simulation in Figure \ref{fig: a}.

\section{SIMULATIONS USING THE SELF-COHERENT CAMERA}
\label{scc}

Although our deterministic speckle nulling procedure in \S\ref{lab} was designed to show that SNWFC is possible experimentally, we would like in the future to further test SNWFC by creating a more realistic dark hole as opposed to nulling a single speckle. In the absence of infrastructure for more stable lab conditions, a higher order DM, and a super-Nyquist phase plate, we instead run simulations to represent the possible performance on a future ELT high contrast imaging instrument.

The SCC, first developed by Baudoz et al. (2006) and evolving most recently to the design and algorithm in Mazoyer et al. (2014), is a wavefront control method to remove speckles in the focal plane\cite{scc1,scc2}. Use of an off-axis hole placed in the Lyot stop causes fringes in the focal plane, but only from the stellar light. Isolating these fringed speckles is done using an amplitude mask on the complex-valued modulus transfer function (MTF), yielding a complex-valued image, $I^-$, which contains phase and amplitude information of only the fringed stellar speckles. By constructing a set reference images in $I^-$ (i.e., an interaction matrix) composed of sines and cosines (each with zero phase) at every $\lambda/D$ interval within a symmetric half-Nyquist region (offset by $0.5\lambda/D$ to preserve symmetry), a least squares fit to an aberrated target image (e.g., with phase and amplitude errors in a sub-Nyquist half-DH region) yields the negative sine and cosine amplitude coefficients to apply to the DM such that the fringed speckles in the focal plane are minimized. See Mazoyer et al. (2014) and references therein for a more detailed description of the SCC wavefront control algorithm.

We choose to use the SCC wavefront control method instead of, e.g., electric field conjugation (EFC)\cite{efc1}, because of the algorithm timescale to converge. It is shown experimentally in in Mazoyer et al. (2014) that the SCC algorithm converges immediately after one iteration\footnote{We note that Mazoyer et al. (2014) do apply a second iteration after the first dark hole is achieved by recalibrating a new interaction matrix, offset with the coefficients already obtained from the initial least squares. We do not perform this second calibration in our simulations, as this is likely done to correct for laboratory limitations that we do not consider in our simulations, such as DM hysteresis, influence functions, fitting error, etc.}, and only relies on the time required to compute the target image correlation vector (i.e., after the initial daytime calibration has been done to compute and invert the interaction matrix). In contrast, EFC, even algorithmically, takes tens of iterations to converge, as in Give'On et al. (2007). With the already calibrated interaction matrix, the SCC timescale to compute the target image correlation vector, sine and cosine amplitude coefficients, and resulting corrected image is $\sim$30 seconds on our 2.7 GHz Mac desktop using one core, but this could ultimately run much faster with a fully parallel code.

The main component of super-Nyquist wavefront control using the SCC, or with any other wavefront control method, requires an optical element that creates a PSF copy outside of the DM Nyquist region. In Thomas et al. (2015), they consider the use of (1) influence functions that are narrow relative to the DM actuator pitch, such as the known manufacturing residual print-through pattern on many DMs, and (2) a mild transmissive diffraction grating (i.e., at a super-Nyquist frequency), and perform SNWFC simulations using option (1). Here, we consider the use of a super-Nyquist sine wave phase plate, initially proposed in Marois et al. (2006) and Sivaramakrishnan et al. (2006) for astrometric and photometric calibration purposes \cite{marois06, sivaramakrishnan}.

We use a  60 nm amplitude sine wave phase plate to create the SCC reference array described above and in Mazoyer et al. (2014), thus copying every sine and cosine from the sub-Nyquist region to the analogous super-Nyquist region. The construction of these reference images and the corresponding least-squares covariance matrix, was done without noise, using 21 nm amplitude sine waves \cite{loci}. This interaction matrix procedure could be implemented on an instrument during daytime calibrations, provided the conditions are stable between daytime calibration and night time observations.
\begin{figure} [!h]
\begin{center}
\includegraphics[height=8cm]{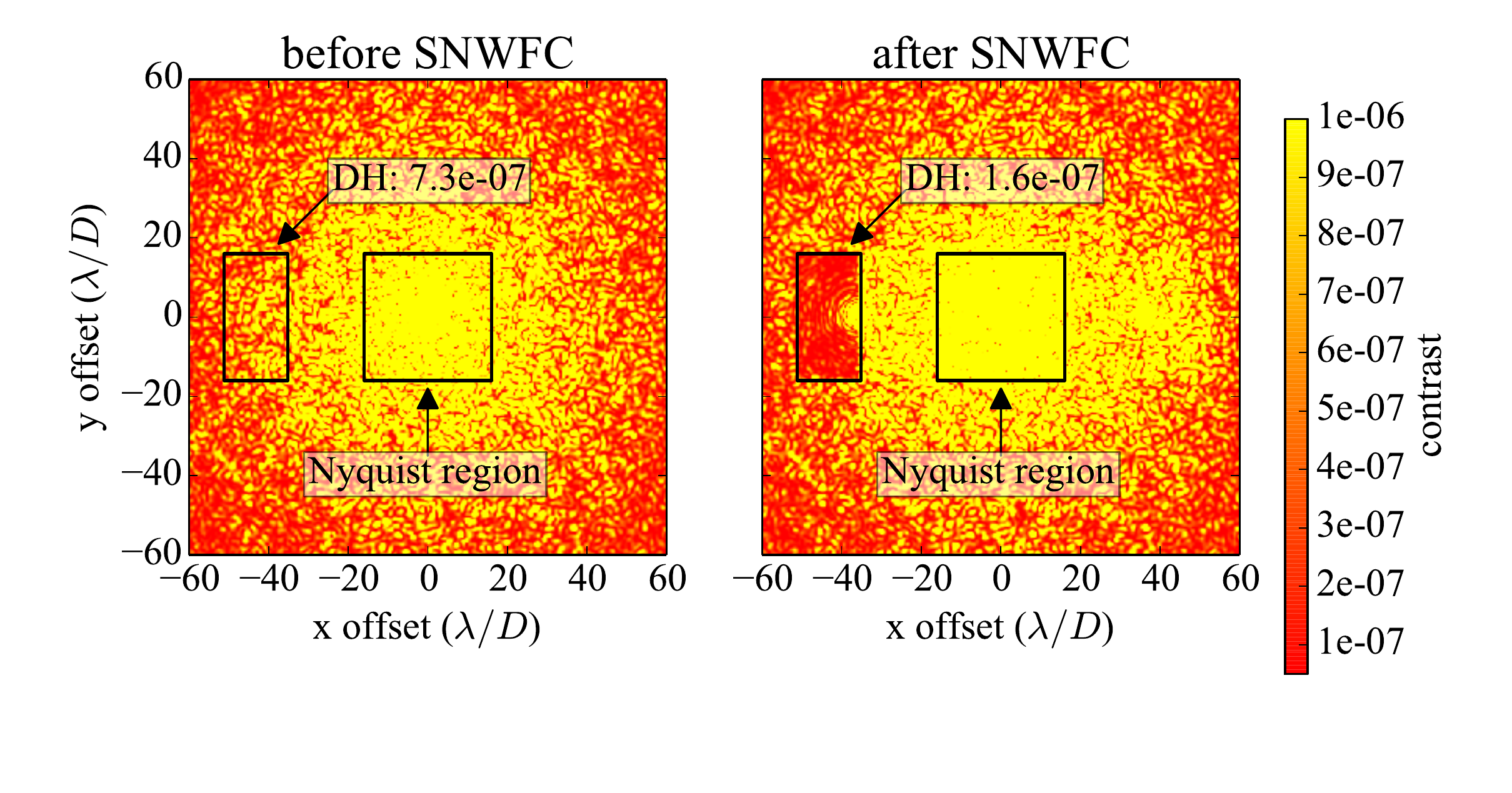}
\end{center}
\caption{ \label{fig: scc} 
Before (left) and after (right) SCC SNWFC simulations, using 40 nm rms wavefront error, 1\% amplitude error, 60 nm amplitude super-Nyquist phase plate (using $\text{cpa}=1.1*N_\text{act}$), a 32$\times$32 DM, and an APLC. The contrast improves by a factor of $\sim$5 after dark hole correction.}
\end{figure} 

Similar to Marois et al. (2012), our target image contained a 40 nm rms phase screen wavefront error and 1\% amplitude errors, both with a -2 power law in spatial frequency\cite{christian_nfiraos}. To simulate the effect of an AO system, we removed the first 21 Zernike polynomials within the DM Nyquist region, using a least squares on the target image with a set of Zernike reference images. We used a 32 by 32 actuator DM, beam ratio of 4 pixels, image plane dimensions of 1012 by 1012, wavelength of 1.65 $\mu$m, and sine wave phase plate cpa of $1.1\left(N_\text{act}\right)$, placing the super-Nyquist region just beyond the sub-Nyquist region. We used the GPI aperture and apodized Lyot coronagraph (APLC) design \cite{remi}. To determine the separation and diameter of the off-axis hole in the Lyot stop that makes the SCC, we used the equations from Mazoyer et al. (2014) ($\epsilon_0>1.5\;D_\text{pupil}$, and $D_R<(1.22\sqrt{2}/N_\text{actuators})D_\text{pupil}$, respectively)\cite{scc2}.

Our simulation results are shown in Figure \ref{fig: scc}. The DH contrast in the raw image is $7.3\times10^{-7}$, and in the SCC image it goes down to $1.6\times10^{-7}$, a gain of $\sim$5 in contrast. We chose a half DH region left of the on-axis PSF, on the left side of the super-Nyquist region, but more generally the phase plate position angle and choice of which side of the spot to build a half DH can be set by the user, if, e.g., the planet location is already known. Images are converted to units of contrast by normalizing the values in the coronagraphic image to the peak value in the initial non-coronagraphic image. To calculate contrast in the super-Nyquist DH-region we take the robust standard deviation\cite{robust} in the contrast-normalized image within an aperture that removes $1\; \lambda/D$ off each edge of the half-DH (to remove edge effects) and $5\;\lambda/D$ radially around the super-Nyquist spot (to remove brighter diffraction effects from the spot that are not present in the raw image without a spot).

\section{CONCLUSION}

SNWFC allows already known and new exoplanets to be directly imaged at a higher SNR than is otherwise possible with the current AO design for ELTs. Our main conclusions in testing this technique are as follows:
\begin{itemize}
\item We have demonstrated in the lab that a deterministic SNWFC speckle nulling scheme increases performance, consistent with simulations.
\item We demonstrated in simulation that SNWFC performance improvement is possible using the SCC after only one iteration, which is of particular interest to wavefront control on shorter timescales for ground-based telescopes.
\end{itemize}

There is still a lot of unexplored parameter space to get a better understanding of how well SCC-based SNWFC would perform on a telescope. Some additional factors in our simulations that are beyond the scope of this initial exploratory paper but worth further research are performance in polychromatic light, using a segmented pupil, using different coronagraphs and/or apodization, dependence on higher super-Nyquist phase plate frequencies, dependence on using an inner and outer scale in chosen wavefront error power law, dependence on a higher zero pad sampling for the off-axis SCC Lyot hole, and additional improvement based on diffraction suppression techniques\cite{efc_l}.  Ultimately, the next step before this technique can be tested on-sky is to demonstrate it in the lab.
\label{conclusion}

\acknowledgments 
 
We gratefully acknowledge research support of the Natural Sciences and Engineering Council (NSERC) of Canada. 

\bibliography{refs} 
\bibliographystyle{spiebib} 

\end{document}